\documentclass{PoS}

\usepackage{siunitx}
\usepackage{floatrow}

\title{An Acoustic Calibration System for the IceCube Upgrade}

\ShortTitle{Acoustic Calibration for the IceCube Upgrade}

\author{
The IceCube Collaboration\footnote{For collaboration list, see PoS(ICRC2019) 1177.}\\
{\itshape \href{http://icecube.wisc.edu/collaboration/authors/icrc19_icecube}{http://icecube.wisc.edu/collaboration/authors/icrc19\_icecube}}\\
E-mail: \email{wiebusch@physik.rwth-aachen.de}
}
\abstract{
The IceCube Neutrino Observatory will be upgraded with about 700 additional optical sensor modules and new calibration devices. Particularly, improved calibration will enhance IceCube's physics capabilities both at low and high neutrino energies. An important ingredient for a good angular resolution of the observatory is a precise calibration of the positions of optical sensors. We present the concept of newly developed acoustic sensors that are mounted inside the optical modules and additional acoustic emitter modules that are attached to the strings. With this system we aim for the calibration of the detectors' geometry with a precision of 10\,cm by means of trilateration of the arrival times of acoustic signals. This new method will allow for an improved and complementary geometry calibration with respect to previously used methods based on optical flashers and drill logging data. 

\vspace{4mm}
{\bfseries Corresponding authors:}
Dirk Heinen$^{1}$,   Shefali$^{1}$, Roxanne Turcotte$^{1}$, Lars Steffen Weinstock$^{1}$,
\speaker{Christopher Wiebusch}$^{1}$, 
Simon Zierke$^{1}$\\
{$^{1}$ \itshape III. Physikalisches Institut B, RWTH Aachen University, D-52056 Aachen, Germany}
}

\FullConference{36th International Cosmic Ray Conference -ICRC2019-\\
		July 24th - August 1st, 2019\\
		Madison, WI, U.S.A.}

\begin{document}

\section{Introduction \label{sec:icecube}}

The IceCube neutrino telescope \cite{Aartsen:2016nxy} is a cubic-kilometer size detector in a depth between 1.5\,km and 2.5\,km in the Antarctic ice at the geographic South Pole. It detects charged secondary particles from neutrino interaction by measuring Cherenkov photons with highly sensitive photo-detectors, the so-called Digital Optical Modules (DOMs).
The  measurement of the arrival time of Cherenkov photons at the sensor positions with ns precision allows reconstruction of the direction and location of particle tracks in the detector. The main uncertainty limiting the accuracy of the directional reconstruction is the optical properties of the medium that affect the photon propagation. In addition, the location of each sensor has to be well known, ideally to an accuracy better  
than $ \SI{1}{ns} \times c_{ice} \simeq \SI{20}{cm}$.
However, within the current calibration scheme of IceCube
%, this goal has not yet been achieved and rather 
an uncertainty of the DOM's position between \SIrange{50}{100}{cm} has to be assumed, depending on depth and the drilling procedure of the specific string. This uncertainty contributes about \SI{10}{\percent} to the total directional uncertainty of  reconstructed  high-energy muon tracks\cite{lili:2019}.

The IceCube Upgrade consists of seven new strings to be deployed near the center of the existing IceCube Neutrino Observatory by 2022/23 \cite{ex13:2019icrc-goal}.
 These upgrade strings will include approximately \num{700} new optical modules, e.g.\ mDOMs \cite{ex2:2019icrc-mdom}, DEggs and others \cite{ex13:2019icrc-goal}, as well as multiple new calibration devices (see e.g.\ \cite{ex1:2019icrc-pocam}, \cite{ex14:2019icrc-pocam}, and \cite{cal3:2019icrc-led}) that will improve  understanding of the ice.
 Most of these new devices will be arranged in a geometry substantially denser than the existing detector with typical string distance of \SI{30}{m} and about \SI{3.5}{m} between DOMs along the strings.
 
One aspect of the improved calibration concept is the integration
of acoustic sensors into the mDOM optical modules. These measure the propagation delay of sound waves that are emitted by seven high-power acoustic pingers that are attached to the detector strings at different depths. Based on the trilateration of  measured propagation times, this new system will lead to an improved and complementary geometry calibration with respect to the previously used methods that are based on the propagation of optical light and on the drill logging data.

Substantial experience has been gathered within the EnEx-RANGE project of the German Space administration, for which a similar system for navigating melting probes in glaciers has been developed \cite{Heinen:2017dih}. 
Additional experience has been gained within the SPATS experiment which deployed acoustic sensors several 100\,m deep into the ice at the South Pole \cite{Abdou:2011cy}.

Main goal of the system is to achieve a spatial resolution of about \SIrange{10}{30}{cm} for the positioning of DOMs. This is similar to the expected resolution by the trilateration of optical signals in the densely instrumented region of the upgrade. This resolution will allow the verification of the optical calibration and the quantification of its systematic uncertainties by an independent system.
%as a function of distance
In particular the uncertainties of the optical propagation delay  due to scattering can be calibrated by comparing the optical and acoustic propagation delay. Based on this,
%correction factors for the 
optical trilateration can be improved 
at larger distances as relevant for the full IceCube detector. When applied to the full IceCube detector, an improved geometry calibration
%of the full IceCube detector 
can be achieved.
This improvement is required to  fully benefit
from future improvements in systematic uncertainties as expected by the upgrade
for reconstruction of events in IceCube.
%expected understanding of photon propagation in the ice in the reconstruction.
%is required to fully exploit the photon information 
%

The acoustic attenuation length is larger than that of optical light in the ice \cite{Abbasi:2010vt}.
Therefore, the acoustic trilateration is a promising method for determining the geometry for even larger string distances of 200\,m and more, as planned for IceCube-Gen2 \cite{Aartsen:2014njl}.
The implementation of  an acoustic calibration system in the current upgrade will thus provide an important proof of method and relevant data for the in-situ performance that are crucial for the design and optimization of the future system for IceCube-Gen2.

Secondary goals of the system are glaciological measurements. This is achieved by detecting
transient acoustic signals related to long-term movement in ice and shear and the re-freezing of holes with coincident observation of transient optical signals \cite{Abbasi:2011zy}, and measuring the speed of sound and its variation versus depth and direction. Furthermore, the system will be capable of searching for acoustic signals in coincidence with optically detected high-energy neutrinos \cite{Abdou:2011cy}.

\section{Design of the Acoustic calibration system}\label{sec:ac-cali}

\subsection{Overall system}

The base design includes three acoustic sensors in every mDOM of the IceCube upgrade. In addition, stand-alone pinger units are positioned close to the locations of the POCAM light sources.
The separation of emitters from the mDOMs is important to avoid potential electromagnetic interference of the high voltage drivers inside the mDOM.
%and allowing for parallel data taking of acoustic and optical systems.
In the absence of  electromagnetic interference, acoustic measurements are optically dark and can be performed continuously and in parallel with the normal operation of the detector.
This will allow operation with a 
low repetition rate and thus low power consumption and will still yield a good signal to noise by averaging a high statistics of transmitted pulses.

\begin{figure}[htp]
% new geometry can be adopted from here https://wiki.icecube.wisc.edu/index.php/Upgrade_Public_Plots#V54_Geometry
	\centering
%\fbox{\color{red} add footprint figure side by side}\\	
	\begin{minipage}[b]{.64\textwidth} \centering
\includegraphics[width=\textwidth]{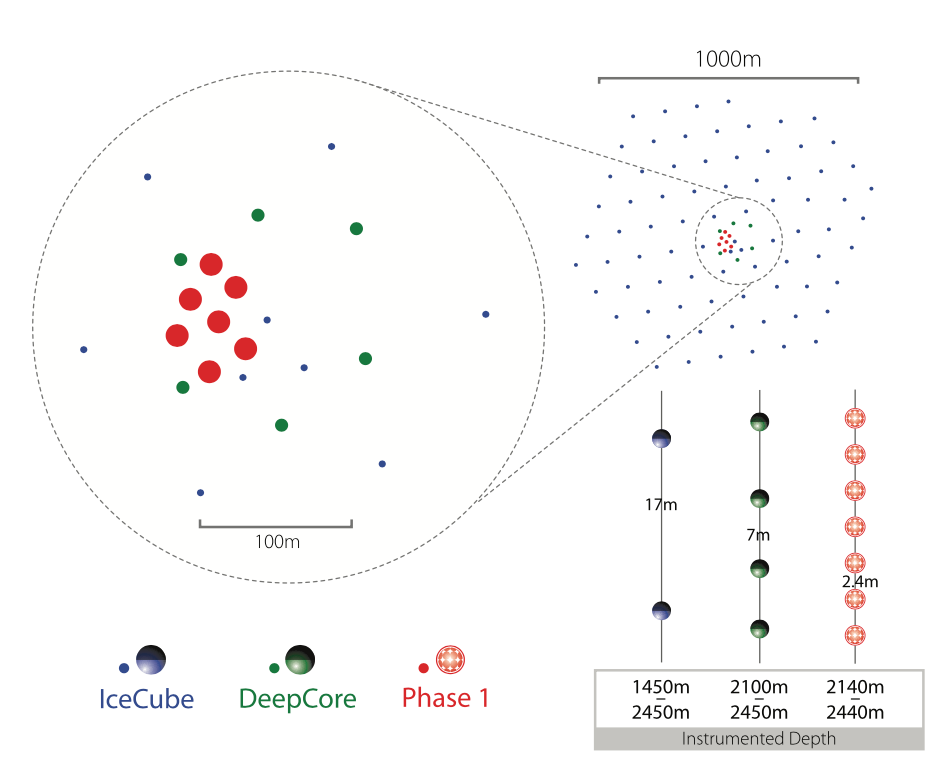} %\\ (a)
\end{minipage} \hfill
\begin{minipage}[b]{.35\textwidth} \centering
\includegraphics[width=\textwidth]{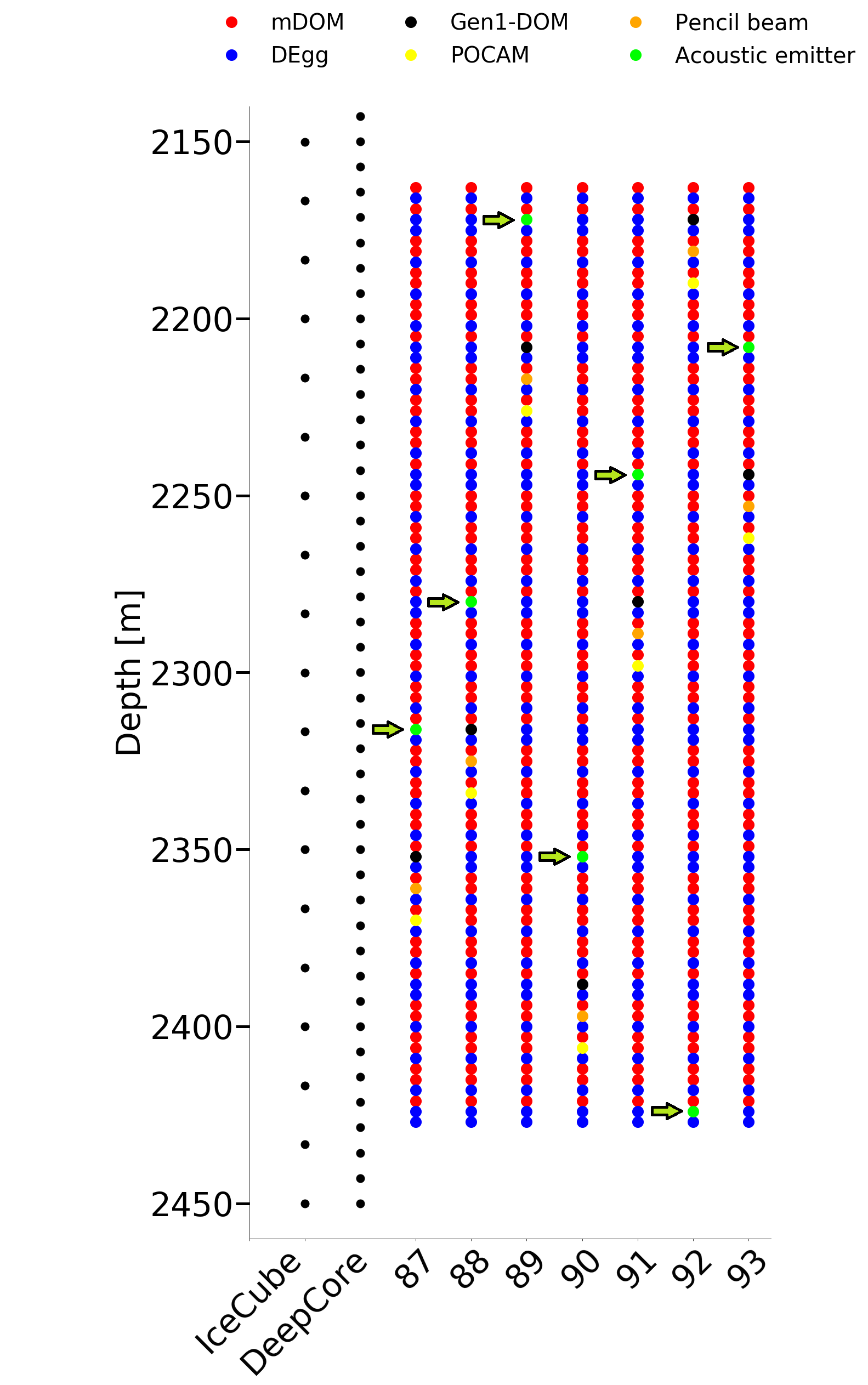} %\\ (b)
\end{minipage}
	\caption{Geometry of the IceCube upgrade. The left figure %(a)  
	shows the footprint of the new strings within the IceCube detector. The densely instrumented region (physics region) between depths of \SIrange{2140}{2440}{m} is shown right %(b) 
	with the positions of the acoustic  pingers marked.}
	\label{fig:geo}
\end{figure}

A tentative geometry with the positions of the strings,
the locations of pingers and mDOMs in the dense instrumented physics region of the upgrade detector is shown in Fig.\ref{fig:geo}. The positions close to optical POCAMs will allow for the direct correlation of optical and acoustic propagation times.

\begin{figure}[htp]
	\centering
%\includegraphics[width=0.46\columnwidth]{Plots/over.png}
	%Regions of positive overconstrainment are indicated by black contour lines that are each separated by \SI{10}{\percent}}
%\includegraphics[width=0.55\columnwidth]{Plots/dof-contours.jpg}
\floatbox[{\capbeside\thisfloatsetup{capbesideposition={right,bottom},capbesidewidth=5.5cm}}]{figure}[\FBwidth]
{\hspace{1cm}\caption{Overconstrainment factor $O$ as function of the number of emitting pingers and receiving mDOMs. Note, that the plot is clipped at zero and white regions correspond to underconstrained combinations. For larger values of receivers and emitters the boundaries remain straight.}}
{\includegraphics[width=6.9cm]{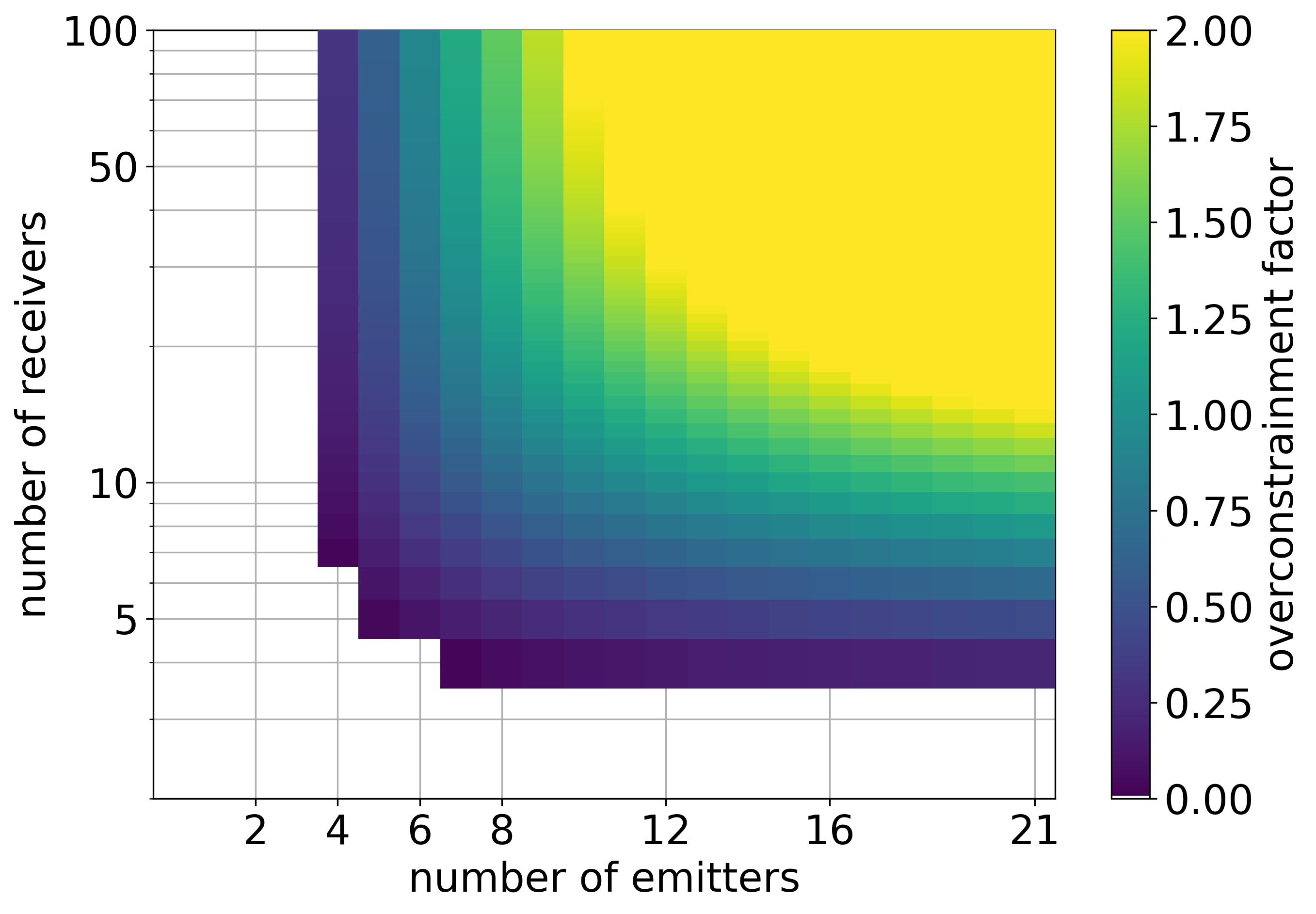}}
	\label{fig:dof}
\end{figure}

For the system design, an important  consideration is the minimum number of emitters and receivers that are required for a good trilateration of the positions.
Given $N$ emitters and $M$ receivers, the number of independent measurements points of transit times is
$n_p = N\cdot M$. 
At the same time, the number of unknown parameters increases with the number of emitters and receivers. Assuming that 
\num{3} coordinates are required for each device and that \num{6} unknowns can be fixed due to the choice of the origin and orientation of an arbitrary coordinate system, the number of unknowns is
$n_u = 3(N+M)-6$. We define the \emph{overconstrainment-factor} $O$ of the trilateration system as 
\begin{equation}
    {O} \equiv \frac{n_p-n_u}{n_u}
\end{equation}
Positive values correspond to a system where sufficient transit times are measured to allow for good trilateration.
The overconstrainment-factor is shown in figure \ref{fig:dof}. At least \numrange{4}{7} pingers and receivers are required for a minimal system. For a system of \num{7} emitters 
that are each well detected by typically 100 receiving modules 
an excellent overconstrainment is reached, with sufficient redundancy against module failures and systematic uncertainties.
Assuming that  not all receivers detect sufficiently good signals from far-away emitters, then the condition $O>0 $ can still be satisfied as long as substantial fractions of the emitter-receiver pairs overlap\footnote{However, in this case the exact form of the above calculation of $n_p $ becomes more complicated.} The uncertainties of the maximum distance scale for good signals will be discussed below. 
However, an acoustic range of the order of \SIrange{50}{100}{m} is fully sufficient to achieve a good overconstrainment.

\subsection{Receivers}\label{sec:receivers}

This receiver concept is based on the acoustic receivers that have been developed for the EnEx \cite{kowalski2016navigation,DissDirk,DissDima} and EnEx-RANGE \cite{Heinen:2017dih} projects.
An  adaption for use in the mDOM was developed and a demonstrator has been 
validated inside a DOM pressure sphere and tested in water \cite{Wickmann:2017zad,turcotte:master:2019}.
Also the compatibility of the optical gel and all sensor materials was tested \cite{turcotte:master:2019}.

\begin{figure}[htp]
%\fbox{\color{red} Add scale for comparison}\\
	\begin{minipage}[b]{.24\textwidth} \centering
	\includegraphics[width=\textwidth]{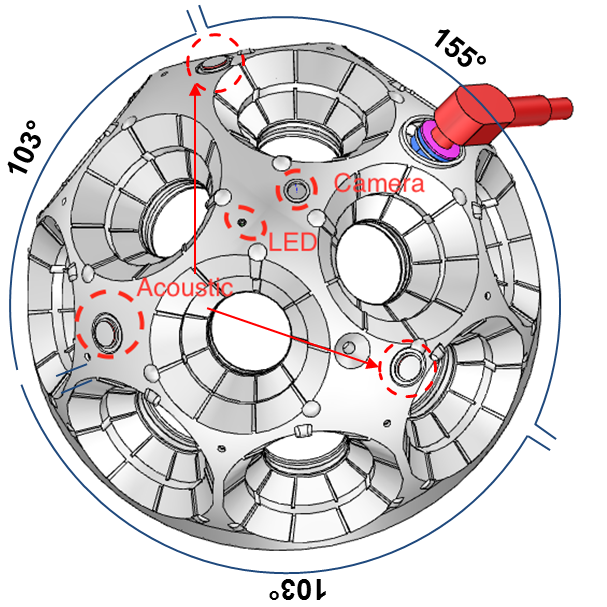} \\
	(a)
	\end{minipage} \hfill
	\begin{minipage}[b]{.18\textwidth} \centering
	\includegraphics[width=\textwidth]{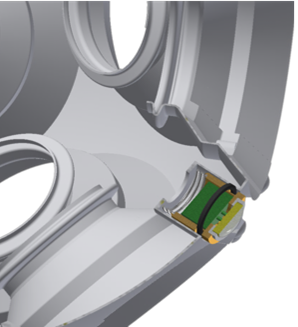} \\
	(b)
	\end{minipage} \hfill
	\begin{minipage}[b]{.3\textwidth} \centering
	\includegraphics[width=\textwidth]{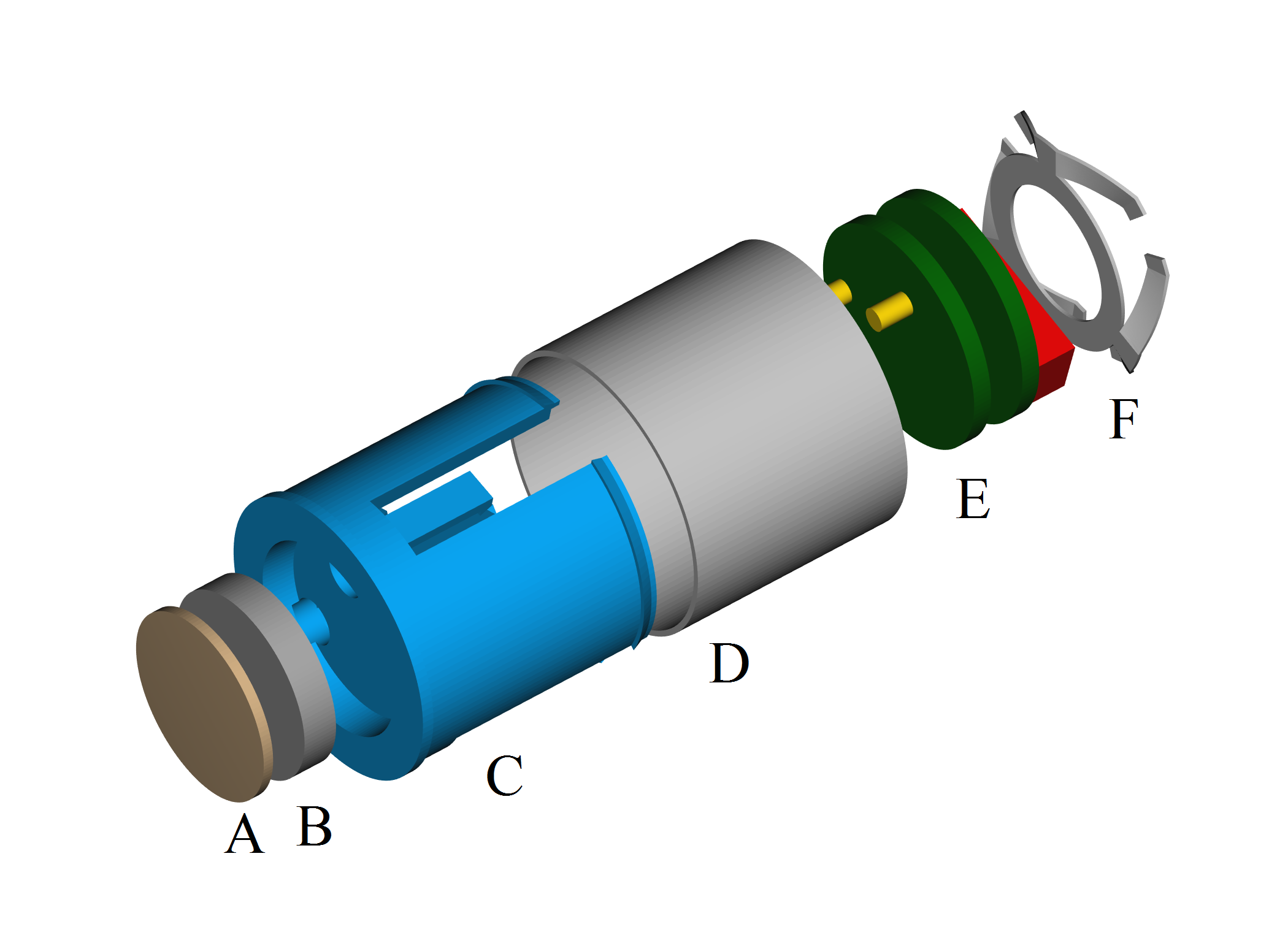} \\
	(c)
	\end{minipage} \hfill
	\begin{minipage}[b]{.17\textwidth} \centering
	\includegraphics[width=\textwidth]{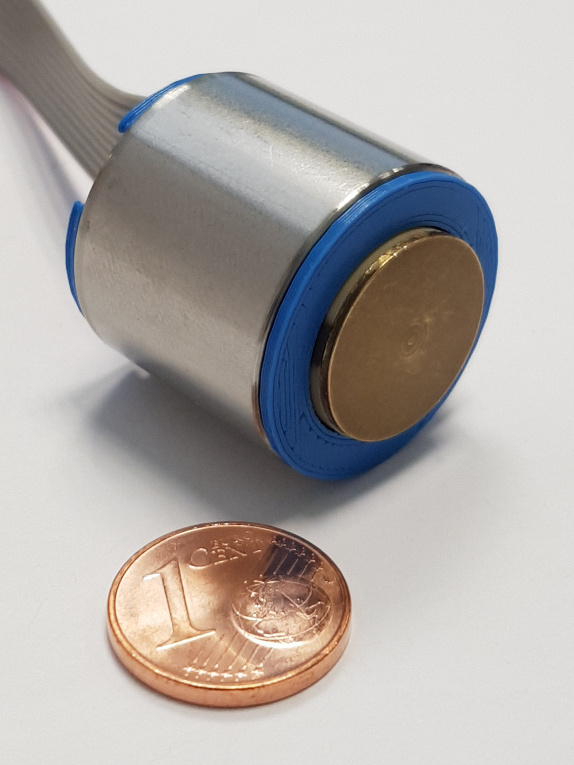}\\
	(d)
		\end{minipage} 
\caption{Acoustic sensor design. The figures show from left to right (a) possible positions within the mDOM's holding structure (b) mounting concept for the sensor in the holding structure (c) exploded view of the sensor and its components (d) picture of an assembled prototype sensor}
	\label{fig:sensor}
\end{figure}

A picture of the sensor design and integration into the mDOM is shown in figure \ref{fig:sensor}.
The mDOM contains a mechanical  holding structure for the photomultipliers to which the sensors are mounted. Three positions, well separated in azimuth direction in order to achieve a good unidirectional response (figure (a)). Note that also an alternative design using only one sensor is under consideration. The sensor will be clipped into the holding structure. The structure as well as the sensor is glued with optical gel into the pressure housing (b). The sensor is sealed with an O-ring to prevent gel from leaking into the inner structure. The mechanical components are illustrated in the exploded view (c).
The transducer (B) is based on a \SI{3}{mm} thick and  \SI{16}{mm} diameter lead zirconate titanate (PZT) disk (type \emph{PIC155} from \emph{PI Ceramic}).
%contact on bounded to th backside
Between the sensor and the glass housing is an optional coupling element made of brass improving the sensors coupling to the spherical glass housing (A). The clip (C) for mounting the sensor into the holding structure is produced by a 3d-printer.
%that is glued to the surrounding pressure housing.
%In addition to this disks the transducer of the pinger (see \ref{sec:pingers}) is used to receive signals.
A metal housing (D) surrounds the sensor and including its electronics (E)  serving as an electromagnetic shielding. An additional spring (F)
ensures a small pressure between the sensor and the housing in order to maintain a good coupling during the curing of the gel.
A fully mounted sensor is shown in (d).
The outer diameter is \SI{24.0}{mm} and the total length is \SI{25.6}{mm}.
% without metal coupler 1#,mm less

%The data handling is done by a FPGA based data acquisition system.
% Three EnEx-sensors were integrated in an IceCube-gen1-DOM and validated in water \cite{Wickmann:2017zad}.

%The current receiver design is shown in figure \ref{fig:sensor}.
%The coupling insert (1) glued to the transducer (2), a PZT disk, sits in a 3d printed plastic structure (3) as well as the front end electronics (5).
%The sensor is electromagnetic shielded by a stainless steel tube (4).
%The steel spring (6) ensures the coupling of the receiver to the glass sphere of the mDOM.
% O-ring and water tightening with epoxy resin ???
%The outer diameter is \SI{24.0}{mm} and the total length is \SI{25.5}{mm}.
% with metal coupler 1#,mm more

%more transducer info? exact material, size, electrodes  

The sensor electronics consists of an analog front-end and the digitization electronics
 implemented as a two stage PCB with an connector in between. All electronics are included in the sensor package, thus making the sensor fully digital in terms of interfacing  to the mDOM main electronics \cite{ex2:2019icrc-mdom}.
 The mDOM mainboard provides power, SPI bus communication and synchronization. 

The front-end electronics 
is directly coupled to the transducer, picking up the signal with two spring loaded pins.
The front end electronics amplify the signal with a variable gain in the range of \SIrange{60}{84}{dB}.
This is achieved by a first stage amplifier of fixed gain and a second stage amplifier of  which gain can be changed by a digital potentiometer.
The amplifier circuit is built as a frequency filter with a pass band of \SIrange{5}{50}{kHz}.

The sensor is controlled by a micro-controller (\emph{Atmel ATSAMD21E18A}), that
handles digitization, local data storage in a \SI{4}{Mbit}  RAM memory,
and communication to the mDOM mainboard  
via a \num{10} wire cable using the  SPI protocol.
The analog to digital conversion is done with variable frequency  \SIrange{100}{200}{kS/s} and  \SI{12}{bit} resolution using an \emph{AD7276} chip.
A local memory of \SI{4}{Mbit} allows the acquisition of up to \SI{2.5}{s} of continuous data.
One sensor total power consumption
is about \SI{50}{mW} and its estimated cost around \SI{50}{Euro}.

\subsection{Pingers}\label{sec:pingers}

\begin{figure}[htp]
	\centering
	\floatbox[{\capbeside\thisfloatsetup{capbesideposition={right,bottom},capbesidewidth=3.5cm}}]{figure}[\FBwidth]
{\hspace{1.cm}\caption{Design of the emitting pinger unit}\label{fig:pinger}}
{\includegraphics[width=10cm,trim={0 2cm 2.5cm 2cm},clip]{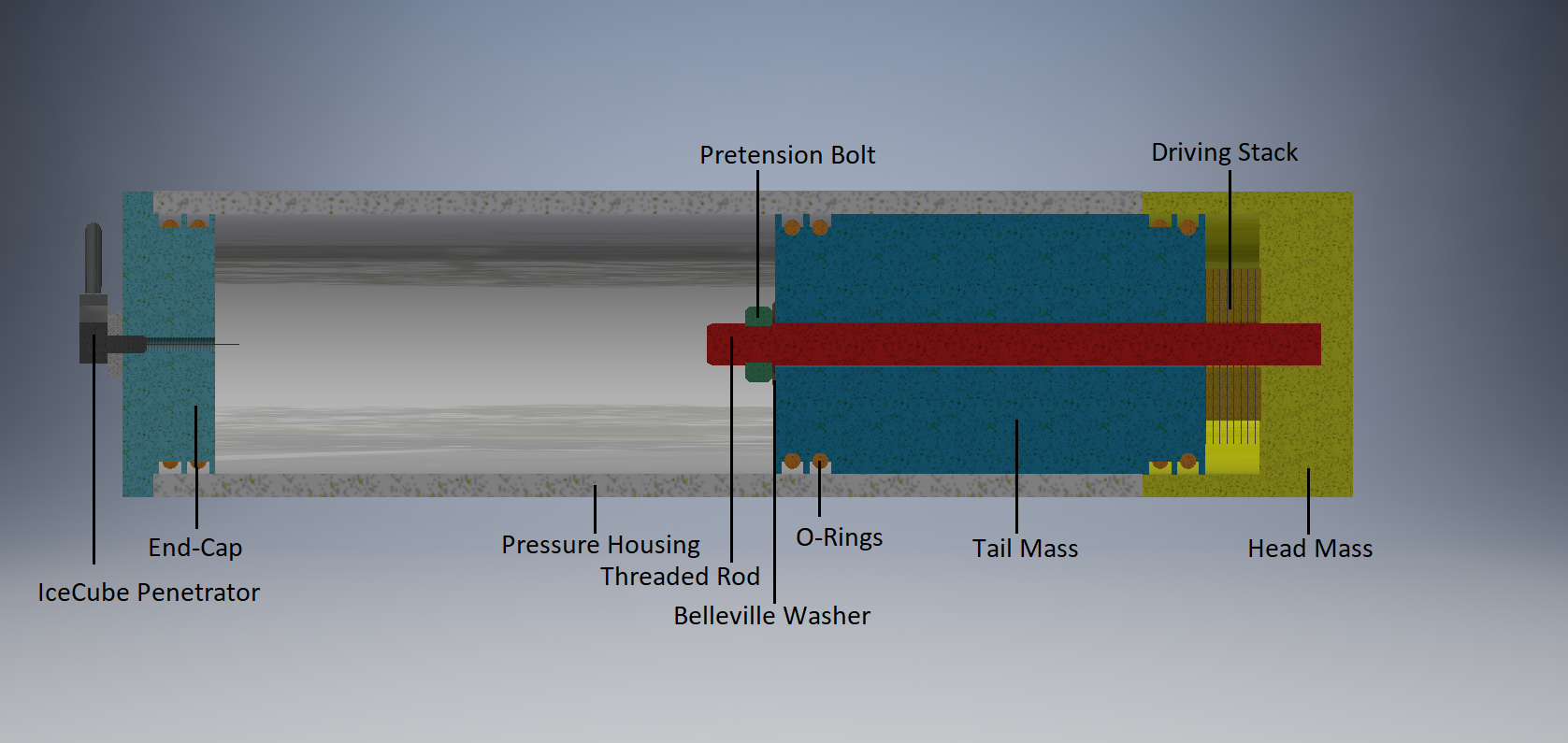}}
\end{figure}

The pingers are standalone devices that are attached to the strings.
They are based on the design that was developed for the EnEx-RANGE project \cite{Heinen:2017dih,DissWeinstock}.
A conceptual drawing is shown in figure \ref{fig:pinger}. The stainless steel housing has about \SI{10}{cm} in diameter and \SI{40}{cm} in length. It holds the mechanical pinger system, the high voltage driving circuits as well as the digital circuit for control and  communication to the surface. A first prototype is currently under construction.

% total weight 14kg
% Konstruktion Pinger
% - Außendurchmesser = 10cm
% - Wandstärke = 1.5cm (Edelstahl, 300bar Außendruck)
% - Länge = 40cm (20cm Platz für Elektronik)
% - Kopfmasse = 2kg, Mittelmasse = 4.5kg, Heckmasse = 9.5kg => Gesamt 16kg
% Elektronik (extrapoliert aus APU)
% - Signal 600Vpp Rechteckspuls (50% dutycycle), Impedanz des Pinger bei Resonanz 200R (Schätzung)
% - 450W Leistung für eine maximale Signallänge von 10ms -> 4.5J elektrische Energie
% - Bei 3.5W Ladeleistung müssen die Speicherkondensatoren etwa 1.3s geladen werden (back of the envelope calculation)

In the center, a stack of ring-like piezo discs (Sonox P4 from CeramTec) of \SI{50}{mm} outer \SI{15}{mm} inner diameter and \SI{2}{mm} thickness is placed. They are arranged in a \emph{Tonpilz} configuration with a front-mass of \SI{2}{kg} and an effective tail mass of \SI{5.5}{kg}. 
The signal is generated by a custom driving circuit that provides waveform signals of frequencies between \SIrange{1}{40}{kHz} implemented as a 3 voltage level signal with voltages of \SIlist{-300;0;+300}{V}.
This allows sending different types of signals, e.g.\ chirps and sine burst, up to \SI{450}{W} of electrical power.
The required energy of a \SI{10}{ms} long signal is then \SI{4.5}{J}.
Assuming that a continuous power of \SI{1}{W}
can be provided to the pinger for recharging, the maximum repetition rates is \SI{0.22}{Hz}. This gives a number of  \num{20e3} pulses per day per pinger that can be measured and averaged for an improved signal-to-noise ratio of the sensor.

\section{Expected performance}

Estimates for the performance of the system are difficult, because they depend on parameters which are not very well known at the moment. While the glass of the sphere matches the acoustic impedance of ice and the piezo material quite well, the mechanical coupling of the sensor to the glass and the coupling of the glass to the ice under high pressure are largely unknown and the directional dependant response function cannot be calibrated prior to deployment. Furthermore, large  uncertainties exist for the propagation of sound in ice. For acoustic trilateration, the most relevant properties are the damping of the signal due to attenuation and the geometric spread with distance, as well as the speed of sound that affects the arrival time. 

The speed of sound can vary moderately with depth and can even 
%vary with horizontal direction
vary within the horizontal plane
in the case where there is a preferential crystal orientation \cite{kluskiewicz2017}. Addressing this uncertainty is only possible with a sufficient overconstrainment factor (see above), that would allow for implementing the speed of sound as an additional fit parameter. Further information can be obtained from external inputs such as ice properties as obtained with optical devices. As the  optical and acoustic trilateration methods are exposed to very different systematic uncertainties from the global ice parameters, their comparison will allow reducing these uncertainties.

The uncertainty of the sensor response within the modules can be tested within water.
Water is an acoustically isotropic and homogeneous medium with negligible attenuation at the considered frequencies.
Several swimming pool tests have been performed with prototype sensors that have been mounted within glass pressure spheres \cite{Wickmann:2017zad,turcotte:master:2019}.
These tests have shown that  the sensors allow for position resolutions of a few \si{cm}.
However, these tests also show that the sensor response within the glass-spheres
has a complex dependency on the direction of the acoustic wave, even if coupling to the medium and the media properties are well known. 
%This has led to the conclusion that 
This is addressed by using
three sensors per mDOM alleviating the problem as this allows for internal consistency checks
%are necessary to allow an internal consistency check
of the measured data such as comparing different sensor orientations with respect to the direction of the incoming acoustic beam.

Trilateration in natural ice and its spatial resolution has been evaluated within the EnEx-RANGE project.
There, the location of a moving melting probe in an Alpine glacier could be determined to an accuracy of typically \SI{30}{cm} \cite{DissZierke}. Taking into account that the positions in IceCube are static, the ice properties of the bulk ice are more favorable (see below), larger averaging factors of repeated wave-forms are possible yielding a better signal-to-noise ratio, and the number of sensors is higher providing  a substantially larger overconstrainment factor, the above value can be considered as robust estimate of the achievable resolution in IceCube.

\begin{figure}[htp]
\centering
\floatbox[{\capbeside\thisfloatsetup{capbesideposition={right,bottom},capbesidewidth=3.5cm}}]{figure}[\FBwidth]
{\hspace{0.5cm}\caption{Expected audible range scaled from EnEx data.}\label{fig:audiblerange}}
{\includegraphics[width=10cm,height=4.5cm]{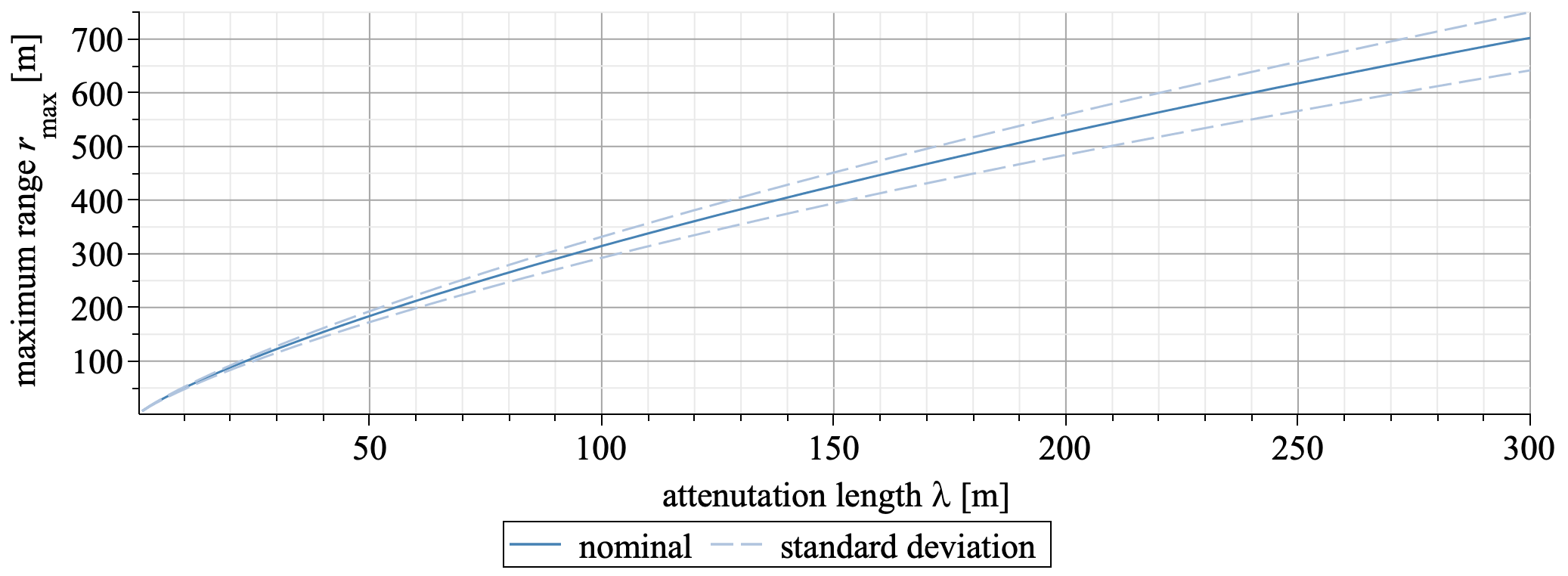}}

\end{figure}

Particularly challenging is the question of acoustic attenuation in the deep ice.
Data and experience from the EnEx-RANGE 
%field tests 
and SPATS can be used 
for an estimation.
Measurements in shallow depth by SPATS  indicate an attenuation length of almost \SI{300}{m} \cite{Abbasi:2010vt}, while EnEx-RANGE measurements \cite{meyer2019attenuation} in a tempered Alpine glacier
give  smaller values of \SIrange{5}{10}{m}. With the ice at larger depth being of similar quality as in SPATS depth but slightly warmer in temperature, we expect an attenuation length substantially larger than for EnEx-RANGE but smaller than measured in SPATS.
Trilateration data from EnEx-RANGE field tests can be used to estimate the acoustic range as a function of the
estimated attenuation length as shown in figure \ref{fig:audiblerange}. The calculation is based on measured 
%signals with distant pingers and good 
signal-to-noise ratios
%for transmitted signals.
%Values up to 
of $S/N = 100 $ %have been 
measured  at distances of $\sim\SI{30}{m}$
with $64$ signal averages %with the EnEx-RANGE System but 
in ice with an attenuation length amplitude of only 
$\lambda = (8.85 \pm 0.95) \si{m} $.
Assuming the damping of the amplitude of a spherical wave with distance $d$ is
\begin{equation}
  A(d) = {a_0 \over d} \cdot \exp{\left (- {d \over \lambda} \right)},   
\end{equation}
the value $a_0$ can be fitted with the measured data. Assuming a minimum required amplitude $A$ (corresponding to the required $S/N$), the
maximum range for a for good transit time measurement can be estimated by
$%\begin{equation}
r_{max} = \lambda \cdot W( {a_0 \over A} \cdot \lambda) )
$%\end{equation}
where $W(x)$ is Lambert W function.

The result of the calculation is shown in figure~\ref{fig:audiblerange}, assuming a minimum $S/N=10:1$, and otherwise the same properties as the EnEx-RANGE system. The shown uncertainty is obtained from the spread of the three best values measured \emph{in-situ}. For a typical attenuation length of 
\SI{100}{m} good transit times will be measurable up to distances of \SI{300}{m}
which is sufficient for the IceCube upgrade, but also for string spacings of the IceCube-Gen2. Note that more averages than the here above assumed \num{64} will further enlarge the usable range.

\section{Conclusion and Outlook\label{sec:conclusion}}

We have presented the design of an acoustic positioning system with the goal of measuring the position of sensors in the IceCube upgrade. The system consists of receivers that are mounted within the optical modules and stand-alone emitters of high power acoustic signals. Initial prototypes of the sensors have been produced and tested and  the emitter prototype is under construction.
Preliminary estimates of the performance are based on previous acoustic trilateration measurements in ice. Those indicate that the system can be scaled to string distances of more than \SI{250}{m} as required for the IceCube-gen2.
With the deployment of the IceCube upgrade, planned in 2022/23,
precise \emph{in situ} data will become available and the design of the system will be further optimized.

%\section{Listing some References}\label{sec:refs}

%Bla

%This is a paper from a previous ICRC %\cite{Zoll:2015wcu}. 
%This is a second paper from a previous ICRC \cite{Peiffer:2017vsm}. 
%This is a paper from the current ICRC \cite{Hussain:2019icrc-gw}.
%Here is an IceCube journal paper \cite{Aartsen:2016nxy} and an external journal paper \cite{Waxman:1998yy}.

% Set up the bibliography using BibTeX.
% Get references from inspirehep.net or NASA/ADS and put them in references.bib.
\bibliographystyle{ICRC}
\bibliography{references}

\providecommand{\href}[2]{#2}\begingroup\raggedright\begin{thebibliography}{10}

\bibitem{Aartsen:2016nxy}
{\bf IceCube} Collaboration, M.~G. Aartsen et~al., {\em JINST} {\bf 12} (2017)
  P03012.

\bibitem{lili:2019}
L.~Peters {Bachelor's Thesis}, RWTH Aachen University, 2019.
\newblock in German.

\bibitem{ex13:2019icrc-goal}
{\bf IceCube} Collaboration,  \pos{PoS(ICRC2019)1031} (these proceedings).

\bibitem{ex2:2019icrc-mdom}
{\bf IceCube} Collaboration,  \pos{PoS(ICRC2019)855} (these proceedings).

\bibitem{ex1:2019icrc-pocam}
{\bf IceCube} Collaboration,  \pos{PoS(ICRC2019)908} (these proceedings).

\bibitem{ex14:2019icrc-pocam}
{\bf IceCube} Collaboration,  \pos{PoS(ICRC2019)928} (these proceedings).

\bibitem{cal3:2019icrc-led}
{\bf IceCube} Collaboration,  \pos{PoS(ICRC2019)923} (these proceedings).

\bibitem{Heinen:2017dih}
D.~Heinen et~al., {\em EPJ Web Conf.} {\bf 135} (2017) 06007.

\bibitem{Abdou:2011cy}
Y.~Abdou et~al., {\em Nucl. Instrum. Meth.} {\bf A683} (2012) 78--90.

\bibitem{Abbasi:2010vt}
{\bf IceCube} Collaboration, R.~Abbasi et~al., {\em Astropart. Phys.} {\bf 34}
  (2011) 382--393.

\bibitem{Aartsen:2014njl}
{\bf IceCube} Collaboration, M.~G. Aartsen et~al.,
  \href{http://arxiv.org/abs/1412.5106}{{\tt arXiv:1412.5106}}.

\bibitem{Abbasi:2011zy}
{\bf IceCube} Collaboration, R.~Abbasi et~al., {\em Astropart. Phys.} {\bf 35}
  (2012) 312--324.

\bibitem{kowalski2016navigation}
J.~Kowalski et~al., {\em Cold Regions Science and Technology} {\bf 123} (2016)
  53--70.

\bibitem{DissDirk}
D.~Heinen.
\newblock PhD thesis, RWTH Aachen University, 2018.
\newblock in German.

\bibitem{DissDima}
D.~Eliseev.
\newblock PhD thesis, RWTH Aachen University, 2018.

\bibitem{Wickmann:2017zad}
S.~Wickmann et~al., {\em EPJ Web Conf.} {\bf 135} (2017) 06003.

\bibitem{turcotte:master:2019}
R.~Turcotte Master's thesis, RWTH Aachen University, 2019.
\newblock internal report icecube/201906001.

\bibitem{DissWeinstock}
L.~S. Weinstock.
\newblock PhD thesis, RWTH Aachen University, 2019.
\newblock in preparation, in German.

\bibitem{kluskiewicz2017}
D.~Kluskiewicz et~al., {\em Journal of Glaciology} {\bf 63} (2017) 603--617.

\bibitem{DissZierke}
S.~Zierke.
\newblock PhD thesis, RWTH Aachen University, 2019.
\newblock submitted, in German.

\bibitem{meyer2019attenuation}
A.~Meyer et~al., {\em The Cryosphere} {\bf 13} (2019) 1381--1394.

\end{thebibliography}\endgroup

% Or, set up the bibliography manually, if you prefer to do things this way.
%
% \begin{thebibliography}{99}
%   \bibitem{Zoll:2015wcu}{{\bf IceCube} Collaboration, \pos{PoS(ICRC2015)1099} (2016).}
%   \bibitem{Peiffer:2017vsm}{{\bf IceCube-Gen2} Collaboration, \pos{PoS(ICRC2017)1052} (2018).}
%   \bibitem{Hussain:2019icrc_gw}{{\bf IceCube} Collaboration, \pos{PoS(ICRC2019)xyz} (these proceedings).}
%   \bibitem{Aartsen:2016nxy}{{\bf IceCube} Collaboration, M.~G.~Aartsen {et al.}, \emph{JINST} {\bf 12} (2017) P03012%
%   % optionally add arXiv ID here [{\tt astro-ph/1612.05093}]
%   .}
%   \bibitem{Waxman:1998yy}{E. Waxman and J. N. Bahcall, \emph{Phys. Rev.} {\bf D59} (1999) 023002.}
% \end{thebibliography}

\end{document}